\documentclass[a4paper,12pt,headings=big,DIV=12]{scrartcl}
\pdfoutput=1
\usepackage{amsmath,amssymb,mathtools}
\usepackage[colorlinks,linkcolor=mygray,urlcolor=blue,citecolor=blue]{hyperref}
\usepackage{tcolorbox} 
\definecolor{mygray}{gray}{0.2}
 \addtokomafont{disposition}{\rmfamily\boldmath}
  \usepackage{bbold}

\definecolor{mypink1}{rgb}{0.9, 0.2, 0.6}
\usepackage{graphicx}
\addtokomafont{disposition}{\rmfamily\boldmath}
\usepackage[utf8]{inputenc}
\usepackage{enumerate} 
\usepackage{slashed}
\usepackage{cite} 
\usepackage{comment}
\usepackage{multirow}
\usepackage{acronym}
 \usepackage{bbold}
\allowdisplaybreaks
\DeclareOldFontCommand{\rm}{\normalfont\rmfamily}{\mathrm}
\DeclareOldFontCommand{\sf}{\normalfont\sffamily}{\mathsf}
\DeclareOldFontCommand{\tt}{\normalfont\ttfamily}{\mathtt}
\DeclareOldFontCommand{\bf}{\normalfont\bfseries}{\mathbf}
\DeclareOldFontCommand{\it}{\normalfont\itshape}{\mathit}
\DeclareOldFontCommand{\sl}{\normalfont\slshape}{\@nomath\sl}
\DeclareOldFontCommand{\sc}{\normalfont\scshape}{\@nomath\sc}

\newcommand{\EQ}{Eq.~}
\newcommand{\EQs}{Eqs.~}

\newcommand{\FIG}{Fig.~}

\newcommand{\SEC}{Sec.~}

\newcommand{\beq}{\begin{equation}}   
\newcommand{\eeq}{\end{equation}}
\newcommand{\beqn}{\begin{eqnarray}}   
\newcommand{\eeqn}{\end{eqnarray}}

\newcommand{\pt}{\partial}

\def\none{${\mathcal N}=1\;$}

\newcommand{\gsim}{\lower.7ex\hbox{$
\;\stackrel{\textstyle>}{\sim}\;$}}
\newcommand{\lsim}{\lower.7ex\hbox{$
\;\stackrel{\textstyle<}{\sim}\;$}}
\newcommand{\vev}[1]{\langle #1 \rangle} 

\newcommand{\al}{\alpha}
\newcommand{\be}{\beta}
\newcommand{\ga}{\gamma}

\newcommand{\de}{\delta}

\newcommand{\la}{\lambda}
\newcommand{\La}{\Lambda}
\newcommand{\eps}{\epsilon}

\newcommand{\sig}{ \sigma}

\newcommand{\calw}{{\mathcal W}}

\newcommand{\ORD}{{\cal O}}

\definecolor{violet}{rgb}{0.94, 0.2, 0.8}
\definecolor{lightblue}{rgb}{0.39, 0.58, 0.93} 
\definecolor{asparagus}{rgb}{0.53, 0.66, 0.42}

\newcommand{\el}{\text{el}}
\newcommand{\ma}{\text{mag}}

\newcommand{\best}{\be_*}
\newcommand{\gastp}{\ga'_{Q^*}}
\newcommand{\bestp}{\be'_*}
\newcommand{\alf}{a_{\rm f}}

\newcommand{\alD}{\al_D}

\newcommand{\Tud}[2]{\theta^{#1}_{\;\; #2}}
\newcommand{\TEMT}{ \Tud{\rho}{\rho} }
\newcommand{\TEMTb}{ \Tud{\al}{\al} }

\newcommand{\lamin}{\la_-}
\newcommand{\lamax}{\la_+}

\newcommand{\CG}{{ \mathcal C}_{G}}
\newcommand{\OG}{{ \mathcal O}_{G}}
\newcommand{\CW}{{ \mathcal C}_{\calw}}
\newcommand{\OW}{{ \mathcal O}_{\calw}}

\usepackage[titles]{tocloft}

\begin{document}

\begin{flushright}
FTPI-MINN-23-19; UMN-TH-4227/23\\
CERN-TH-2023-188 \\
\today
\end{flushright}

\vspace{3mm}

\begin{center}
{\large\bfseries \boldmath Relating   $\beta'_*$ and $\gamma'_{Q*}$ 
 in  the ${\cal N}=1$ SQCD Conformal Window }
\\[0.8 cm]
{\large%
 Mikhail Shifman$^{a}$ and Roman Zwicky$^{b,c}$
\\[0.5 cm]
\small 
$^a$ William I. Fine Theoretical Physics Institute, \\
University of Minnesota, Minneapolis, MN 55455 \\[0.2cm]
 $^b$ Higgs Centre for Theoretical Physics, School of Physics and
Astronomy, \\ The University of Edinburgh  \\[0.2cm]
$^c$ Theoretical Physics Department, CERN, \\
Esplanade des Particules 1,  Geneva CH-1211, Switzerland
} \\[0.5 cm]
\small
E-Mail:
\texttt{\href{mailto:j.shifman@umn.edu}{shifman@umn.edu,}
\href{mailto:roman.zwicky@ed.ac.uk}{roman.zwicky@ed.ac.uk}}
\end{center}

\bigskip
\thispagestyle{empty}

\begin{abstract}\noindent In this note we  show that $\beta'_*$, the $\beta$-function slopes in the electric and magnetic theories
are equal at the corresponding infrared fixed points.
This follows from the scaling of the correlators of the trace of the energy momentum tensors.
The slopes $\beta'_*$  determine the scaling dimensions. 
Our paper can be considered as a commentary to Anselmi et al. \cite{AGJ} -- it proposes an improved derivation 
not based on a rather contrived construction by Kutasov et al. \cite{Kutasov}. 
As a byproduct we note that $\gamma'_{Q^*}$ -- the slopes of the matter superfield anomalous dimension --
 vanish at both edges of the conformal window where one of the dual theories is strongly coupled. Finally,
we determine the two-coupling  magnetic fixed point at weak coupling correcting the result of \cite{KSV}.
 \end{abstract}


%
%
%

\newpage

\setcounter{page}{1}

\mdseries 

\tableofcontents

\section{Introduction}
\label{intro}

Yang-Mills theories with ${\cal N}=1$ supersymmetry 
 produce a wide  variety of exact 
results. One of the most important is the Seiberg duality (for reviews see \cite{Intriligator:1995au},\cite{Shifman:1999mk},\cite{Terning:2006bq},\cite{Tachikawa:2018sae}) which states that in 
 deep infrared (IR) an ${\rm SU}(N_c)$ gauge theory of $N_f$ flavors  is dual to a ${\rm SU}(N_D)$ ($N_D \equiv N_f-N_c$) gauge theory with $N_f$ flavors   and $N_f^2$ color-singlet mesons.  
As is usual with dualities, when the original (electric) theory is weakly 
coupled then the dual theory, referred to as magnetic, is strongly coupled, and vice versa.  

In the conformal window (CW) which lies in the interval
\beq
\frac 32 N_c \leq N_f \leq 3 N_c \;,
\label{one}
\eeq
both theories are asymptotically free and conformal in the IR due to an IR fixed point. We will limit ourselves to the window 
(\ref{one}) assuming that $N_{f,c}\gg 1$ with $N_f/N_c$ fixed.
The lower and upper boundaries  are referred to as the CW edges. 
Above the upper edge the electric theory becomes free while at $N_f< \tfrac{3}{2} N_c$  the same transition happens in the magnetic theory. The edges can be obtained either
from the NSVZ beta functions \cite{Novikov:1983uc,Novikov:1985rd} or from the unitarity bound \cite{Seiberg:1994pq}.
Recently an alternative interpretation has been given in terms of a smooth matching  
to the chirally broken phase with pion physics \cite{Zwicky:2023bzk}.

In this paper we will focus on the slopes of the $\beta$ functions in both dual theories in the deep IR (labelled by the subscript $_*$),
\beq
\beta_*^\prime = \left.\frac{\partial}{\partial\alpha} \beta \right|_*\,.
\eeq
Since the above slope is related to the scaling dimension of a physically observable operator, the trace of the energy-momentum tensor (TEMT), 
which has a geometrical meaning, the slopes in the electric and magnetic theories must coincide,
\beq
 \left. \beta_*^\prime\right|_{\rm el} =  \left. \beta_*^\prime\right|_{\rm mag}\,,
 \label{three}
 \eeq
 for each given value of $N_f$ from the CW.  For a
 the definition of $\beta^\prime_*|_{\rm mag}$  see \SEC\ref{sec:mag}.
 
 Equation (\ref{three}) was originally obtained in \cite{AGJ} by analyzing the Konishi currents  in both dual theories on the basis of the 
 Kutasov construction \cite{Kutasov}.\footnote{That \EQ \eqref{three}  could be true 
 is supported indirectly by the fact that in perturbation theory the sign of $\be'|_{\el}$ alternatives at 
 higher orders in the expansion \cite{Ryttov:2017kmx}.}
 Our goal in this paper is to bypass the Kutasov construction which is not needed for the derivation of (\ref{three}).
 Also, we derive a previously unknown relation between {$\gastp$}  and  $\bestp $ in the electric theory and revisit numerical calculations
 near the CW edges.

The organization of the paper is as follows.
 In \SEC\ref{prel} we collect some basic elements of the \none theories which are dual in the CW. 
\SEC\ref{sec:electric}
is devoted to the study of the two-point function of the TEMTs which allows us to establish the anomalous dimension of the $x$
dependence of the   two-point function in question In \SEC\ref{sec:mag} we address 
the analogous  two-point function in the magnetic theory.
In the latter, in addition to the gauge interaction, a Yukawa interaction is present too, proportional to a meson and two quark superfields.  
Therefore, instead of the single $\beta$ function of the electric theory, in the magnetic theory  we have to deal with two $\beta$ functions.
We define the notion of $\beta^\prime_*$ in the magnetic theory and determine this quantity. In \SEC\ref{sec5} we derive a new relation between
$\gastp$  and  $\bestp $. In \SEC\ref{edgess} we calculate  $\beta^\prime$s and other necessary parameters at weak coupling near
the edges of the CW. We also determine a stable IR fixed point which corrects the result of \cite{KSV}. Finally, the Appendix is devoted to
the derivation of the TEMT in the supersymmetric formalism and the $R$ and Konishi current correlators.


\section{Preliminaries and notation}
\label{prel}

In this section we outline the formalism to be used below and introduce our notation. The latter follows the second book in \cite{Shifman:1999mk} (which, in turn, is very close to that of Wess and Bagger  \cite{WB}).

The electric theory contains {\rm SU}($N_c$) gauge bosons and the following matter sector: $2N_f$ chiral superfields in the fundamental representation,
namely, $N_f$ findamentals $Q_k$ and $N_f$ anti-findamentals $\tilde{Q}^k$,  $k=1,2,...,N_f$. The Lagrangian has the form
\beqn
\mathcal{L}&=& \left(\displaystyle \frac{1}{4g^{2}}\int d^{2}\theta
W^{a\alpha}\, W_{\alpha}^{a}+\mathrm{H.c.}\right)+\sum_{\mathrm{all\
flavors}}\left\{\int\left(
d^{2}\theta\,d^2\bar{\theta}\, \bar{Q}^{\,\bar f}e^{V}Q_{f}\right.\right.\nonumber\\
&+&\left. \int \left.
d^{2}\theta\,d^2\bar{\theta}\, \bar{\tilde{Q}}^{\,\bar f}e^{-V}\tilde Q_{f}\right)\right\}.
\label{four}
\eeqn
The above Lagrangian is written in the ultraviolet (UV); as we descend down to the IR,  $1/g^2$ is replaced by the runnnig constant $1/g(\mu)^2$
and the matter-field Z factor ($Z_Q(\mu)$)  appears in front of the second term in (\ref{four}).\footnote{ We omit the subscript el where there is no menace of confusion.}

In the dual magnetic theory the dual quark superfields are denoted as $q_f$ and $\tilde {q}^f$
and the dual color is 
\beq
\label{five}
N_D \equiv N_f-N_c\,.
\eeq
In addition, in the magnetic theory  one has to introduce a color singlet matter-field represented by the matrix $ M^i_{{j}}$ and the superpotential 
\beq
{\cal W} = {\rm f}\, M^i_{{j}} q_i \tilde{q}^{\, j} \;,
\label{eqn14}
\eeq
where $\rm f$ is the Yukawa coupling which can be chosen to be real. The corresponding kinetic term is normalized canonically, ${\rm Tr} \!\left( \bar{M} M \right) $.

{The $\beta$ function and the matter anomalous dimension are defined as}
\beqn
\beta(\alpha )&=&\frac{\partial \alpha (\mu) }{\partial L} \,,\quad \alpha = \frac{g^2}{4\pi},
\\[2mm]
\ga_Q &=& -\frac{d\log Z_Q}{dL}\,, \quad L=\log\mu\,.\nonumber
\eeqn
In the dual magnetic theory we will introduce $g^2_D$ and $\alpha_D = \frac{g^2_D}{4\pi}$.

Finally we will need the expression for the NSVZ beta function \cite{Novikov:1983uc,Novikov:1985rd}, 
\beq
\beta =-\frac{\alpha^{2}}{2\pi}
\big[3N_c-N_f(1-\gamma_Q)\big]\left(1-\frac{N_c\,\alpha}{2\pi}\right)^{-1}.
\label{eight}
\eeq
At one loop 
\beq
\gamma_Q =-\frac{\alpha}{\pi} \frac{N_c^2-1}{2N_c}\to -\frac{\alpha}{2\pi}N_c \,.
\label{9}
\eeq

We will also need the expression for the hypercurrent divergence which includes all three geometric anomalies. 
In the operator form it can be read off from Eq. (\ref{eqn49}) provided we omit the first line which is needed only in the magnetic theory.
Then we conclude that 
\beq
\left(\theta^\rho_{\,\rho}\right)_{\rm el} = \CG \, \beta(\alpha) \, \OG  \;,
\label{9p}
\eeq
where 
\beq
\OG   = - 2 {\rm Re} \left[W^{\alpha\,a} W^a_\alpha 
\right]_{\theta^2{\rm el} } =
\left[G_{\mu\nu}^a G^{\mu\nu\, a}-2D^2 -4i\bar{\lambda}_{\dot\alpha}^a {\mathcal D}^{\dot\alpha\alpha}\lambda_{\alpha}^a
\right]_{\rm el} \;,
\label{10p}
\eeq
and
\beq
\CG =  \left(16\pi\,\alpha^2 \right)^{-1} \,.
\label{11p}
\eeq
Here and in what follows we will refer to $\OG$ in the right-hand side of (\ref{10p}) as $G^2$; hence, the corresponding notation.

\section{  \boldmath{$\langle \theta^\rho_\rho (x) \theta^\alpha_\alpha (0)\rangle $} in the electric theory}
\label{sec:electric}

For a generic local operator $O(x) $ 
the two-point function $\langle O(x) O(0)\rangle$ in the conformal limit takes the form $(x^2)^{-\Delta_O}$
where $\Delta_O$ is the sum of the normal (engineering) and anomalous dimensions of the operator $O$,
\beq
\Delta_{O}=d_{O}+\gamma_{O} \;.
\eeq
The engineering dimension of TEMT  obviously is $d_{G^2} =4$.  We will focus on the anomalous dimension.

The energy-momentum tensor $\theta_{\al\be}$ 
is symmetric and conserved and has a geometric nature.
This tensor itself does not change  under the variation of $\mu$, as we descend from the UV to the IR.
Hence, its trace is not renormalized either. The particular expression for $ \theta^\rho_\rho$ depends on how we normalize 
the gluon field stress tensor, but the final result for the $x$ scaling  dependence near the conformal point
is unabiguously determined by the theory and is physical,
\beq
\langle \OG (x)   \OG (0) \rangle \propto \frac{1}{(x^2)^{\Delta_{G^2}}}\,,\qquad  \Delta_{G^2}=d_{G^2}+\gamma_{G^2} \,,
\label{nine}
\eeq
where  $d_{G^2}$ is the engineering dimension of the operator $\OG$ and  $\gamma_{G^2}$ is the anomalous dimension of $\OG$.

Without loss of generality we can choose $\OG$ normalized in accordance with \EQ(\ref{four}). The TEMT is given in (\ref{9p})-(\ref{11p})
where the $\beta$ function in the electric theory is presented in (\ref{eight}). Differentiating both sides over $L=\log\mu$ we arrive at
\beq
0= 
\OG \,  \frac{\partial }{\partial L} \left(\frac{\beta}{\alpha^2}\right) +\frac{\beta}{\alpha^2}  \frac{\partial \OG }{\partial L} \,,
 \label{eleven}
\eeq
which implies, in turn, 
\beq
\beta(\alpha)  \OG  \frac{\partial}{\partial\alpha}  \left(\frac{\beta}{\alpha^2}\right)- \gamma_{G^2} \, \OG  \frac{\beta}{\alpha^2}=0\,.
 \label{twelve}
\eeq
Taking into account that $\beta (\alpha_*)=0$ while $\alpha_*\neq 0$ we find the well-known result 
\beq
\left(\gamma_{G^2}\right)_* = \left.\frac{\partial\beta}{\partial\alpha}\right|_*\equiv  \beta^\prime_*\,.
\label{thirteen}
\eeq

\vspace{2mm}

Finally we can present the two-point function of the TEMTs at the points $x$ and $0$. Note that the $x$ dependence is fully determined by the 
two-point function (\ref{nine}), therefore,

\beqn
\langle \theta^\rho_\rho (x) \theta^\rho_\rho (0) \rangle &\propto & \left[\left.\left(16\pi\,\alpha^2 \right)^{-1} \beta(\alpha)\right|_\mu\right]^2 \langle  \OG (x)\, \OG (0)
\rangle_\mu\nonumber\\[2mm]
&\propto &\left[\left.\left(16\pi\,\alpha^2 \right)^{-1} \beta(\alpha)\right|_\mu\right]^2\,\frac{1}{(x^2)^4} \, \frac{1}{(x^2\mu^2)^{\beta^\prime_*}} \;.
\label{fourteen}
\eeqn
The $\mu$ dependence in Eq. (\ref{fourteen}) enters explicitly through $\mu^{-2\beta_*^\prime}$, and implicitly, through the prefactor.
{At the conformal point $\mu\to 0$ the scale dependence of the prefactor 
$$
P= \left.\left(16\pi\,\alpha^2 \right)^{-1} \beta(\alpha)\right|_\mu\,.
$$
is}
\beq
P \to \left( \frac{\mu}{\Lambda}\right)^{\beta_*^\prime} \,,
\label{15p}
\eeq
where $\Lambda$ is a $\mu$ independent  scale parameter which can be seen as the analogue of 
$\La_{\text{QCD}}$ that determines the logarithmic running in the chirally broken phase. 
As a result, we arrive at the following final result for the correlator at hand,
\beq
\langle \theta^\rho_\rho (x) \theta^\alpha_\alpha (0) \rangle \propto \frac{1}{(x^2)^4} \, \frac{1}{(x^2\Lambda^2)^{\beta^\prime_*}}\,.
\label{16p}
\eeq
This is our main result in this section. 
For completeness, and since it follows rather directly, we 
 derive the $R$  and Konishi current correlators in  Appendix \ref{app:KR}.

\vspace{1mm}

\section{ \boldmath{$\langle \theta^\rho_\rho (x) \theta^\alpha_\alpha (0)\rangle $} in the magnetic theory}
\label{sec:mag}

In the magnetic theory  the  number of colors $N_D$ is given in (\ref{five}).
In other words,  SU$_{\rm gauge}(N_c)\to$ SU$_{\rm gauge}(N_f-N_c)$, with the same CW
\beq
\frac 3 2 N_D \leq N_f\leq 3N_D \,.
\eeq
The  matter fields of the magnetic theory  $q_i$ and $\tilde{q}^{j}$ belong to the (anti)fundamental representations
of ${\rm SU}(N_D$). In addition one  must add a color-singlet matrix field $M_j^i$, depending on the flavor indices, and the superpotential shown in Eq. (\ref{eqn14})
where $ {\rm f}$ is the ``second" holomorphic coupling constant of the theory. The Lagrangian takes the form
\beqn
\mathcal{L}&=& \left(\displaystyle \frac{1}{4g_D^{2}(\mu)}\int d^{2}\theta
W^{a\alpha}W_{\alpha}^{a}+\mathrm{H.c.}\right)+\sum_{\mathrm{all\
flavors}}Z_q(\mu)\int
d^{2}\theta\,d^2\bar{\theta}\, \tilde{q}^{\,\bar f}e^{V}q_{f}\nonumber\\
&+& \int
d^{2}\theta\,d^2\bar{\theta}\left[ Z_M(\mu) \,{\rm Tr}\, \left( \bar{M} M \right) \right]    + \left(\int d^{2}\theta\,
\mathcal{W}(\tilde{q}, q,M)+\mathrm{H.c.}\right) \;,
\label{eqn15}
\eeqn
{where the superpotential  $\mathcal{W}$ is defined in \eqref{eqn14}.}
The $M$ superfield is in the magnetic representation; therefore its dimension in the UV is one, and the coupling $ {\rm f}$ is dimensionless.
The trace in the second line of (\ref{eqn15}) runs over flavors. Lagrangian (\ref{eqn15}) explicitly exhibits the effect of the RG flow. {At the classical level it is scale and conformal invariant.}
The $\mu$ dependence breaks the scale symmetry and gives rise to two $\beta$ functions,
$\beta_{D}$ for the gauge coupling and $\beta_{\rm f}$ for the super-Yukawa coupling. The latter appears only due to  the $Z_{q,M}$ factors when we apply the equations of motion. Indeed, in passing to the canonically normalized matter kinetic terms we obtain
\beq
\frac{ {\rm f}^2}{4\pi} \to  \frac{{\rm f}(\mu)^2}{4\pi} = \frac{ {\rm f^2}}{4\pi} \left[Z_q(\mu)^2{Z_M(\mu)}\right]^{-1}\,.
\label{27}
\eeq
In what follows we will use the notation
\beq
\alf \equiv \frac{ {\rm f}^2}{4\pi}\,.
\eeq

Then the expressions for the magnetic $\beta$ functions are as follows
\beqn
\label{eq:beM}
\beta_{\rm f} (\alpha_D, \,\alf)&\equiv& \frac{\pt}{\pt L}a_{\rm f} = 
 a_{\rm f} \left[ \gamma_M (\alpha_D, \,\alf)+2\gamma_q(\alpha_D, \,\alf) \right]\,,\nonumber\\[2mm]
\beta_{ D} (\alpha_D,  a_{\rm f}) &=&  -\frac{\alpha^{2}_D}{2\pi}
\big[3N_D-N_f(1-\gamma_q(\alD, {\rm f}))\big]\left(1-\frac{N_D\,\alpha_D}{2\pi}\right)^{-1} \;,
\eeqn
where $\beta_{ D}$ is the  $\be$ function \eqref{eight} with $N_c \to N_D$ and $\ga_Q \to \ga_q$  
and   the one for $\be_{\rm f}$ is obtained by differentiating \eqref{27}.  
The latter  holds in principle up to non-perturbative corrections since the non-renormalization theorem of 
the superpotential is perturbative in nature.  However at weak coupling non-perturbative corrections are exhausted 
by instantons but they are absent since the $R$-symmetry implies that the zero modes do not match for 
$N_f > N_c +1$.  Moreover, it has been argued that the absence of renormalon ambiguities inside 
the conformal window implies the absence of non-perturbative corrections \cite{MS}. We therefore 
conjecture that the expression for $\beta_{\rm f}$ is formally correct  for $N_f > N_c +1$.  
 
At the critical values of the coupling constants the sum of anomalous dimensions vanishes,
$$
\gamma_{M*} + 2 \gamma_{q*} = 0\,.
$$ 
Explicit leading order expressions are given in \SEC\ref{edgess}.



In what follow we aim to show that  \eqref{16p} holds equally for the magnetic theory
when suitably adapted.  
This requires a bit more work since  in the magnetic theory we have two couplings. 
The TEMT in the magnetic theory, given in  \eqref{eq:TEMT},  reads 
\begin{equation}
\label{eq:TEMTmag}
\TEMT =   \CG \, \be_D(\alpha_D, \,\alf) \, \OG +     \CW \, \beta_{\rm f} (\alpha_D, \,\alf) \, \OW  \;, 
\end{equation}
where  $\OG$  and $\OW$ corresponds to the gluon and superpotential part respectively, see \EQs(\ref{a3})-(\ref{a5}) in the Appendix \ref{app:defs}.
We may linearize  both $\be$ functions around the IR fixed point.  If we
define the coupling vector 
\beq
\de \underline{\al} \equiv \left( \begin{array} {l}
{\alpha_D} - {\alpha_{D_*}} \\[2mm] a_{\rm f}-a_{{\rm f}_*}
\end{array}
\right) \;,
\label{32}
\eeq
the linearized $\be$ function can be written as
\begin{equation}
\label{eq:be2}
\frac{\partial}{\partial L}  \de \underline{\al} = B_*  \, \de \underline{\al}  + \ORD((\de \underline{\al})^2) \;, 
\end{equation}
where  $B_*$ is the gradient matrix 
\begin{equation}
\label{eq:L}
B_* =  \left(  \begin{array}{ll}  
 \partial_{\alD} \be_{D} &  \partial_{\alf} \be_{D}   \\[2mm]  
  \partial_{\alD} \be_{f}   &   \partial_{\alf} \be_{f} \end{array} \right)_* \;,
\end{equation}
evaluated at the IR fixed point. The derivatives in the matrix $B$ are just numbers independent  of the running $\alpha_D$ and $a_{\rm f}$
which can depend, however on the numerical values of ${\alpha_{D_*}}$  and $a_{{\rm f}_*}$ (cf. \SEC\ref{edgess}).

Next, we can diagonalize  the matrix $B_*$, which  generically has two unequal
real eigenvalues $\lamin  \leq \lamax $.\footnote{\label{foot:Wein} These eigenvalues are
 scheme-independent  under analytic coupling redefinitions e.g. \cite{Weinberg:1996kr}.}
Indeed, let us introduce the matrix $U$ diagonalizing $B_*$,
\beq
B_{\rm diag} \overset{\rm def}{=}\hat{B} =U^{-1}B_*U\,,\qquad \hat{B} =\left(\begin{array}{cc} \lambda_-&0\\[2mm] 0&\lambda_+
\end{array}\right).
\label{35}
\eeq
Correspondingly, the ``diagonalized form" for the column $\de \underline{\al}$ in (\ref{32}) becomes
\beq
\de \underline{\hat\al}  \equiv  { U}^{-1}  \de \underline{\al}\,.
\eeq
Then we can write
\beq
\frac{d}{dL} \de \underline{\al}= B_*  \, \de \underline{\al}= U\hat{B} \de \underline{\hat\al} \;,
\label{37}
\eeq
or, alternatively
\beq
\frac{d}{dL} \de \underline{\hat\al}=\hat{B} \de \underline{\hat\al}\,.
\label{38}
\eeq
The solution of the equation above  takes the form, 
\beq
\de \underline{\hat\al}=\left(\begin{array}{c}\left(\frac{\mu_{\phantom{-}}}{\La_-}\right)^{\lambda_-} \\[1mm]
\left(\frac{\mu_{\phantom{+}}}{\La_+}\right)^{\lambda_+}
\end{array}
\right) \;,
\label{39}
\eeq
where the eigenvalues $\lambda_\mp$ have to be non-negative and $\La_{\mp}$ are related to the choice of initial condition, 
cf. \FIG\ref{fig:flowmag}.

Now let us return to Eq. (\ref{eq:TEMTmag}). Introducing a row of operators $\cal O$,
\beq
{\cal O }= \left\{  \CG\, \OG, \CW \, \OW \right\} \;,
\eeq
we can rewrite  (\ref{eq:TEMTmag}),  using  the linear approximation in Eq. (\ref{37}), as follows,
\beq
\theta^\rho_{\,\rho} = {\cal O} B_*  \, \de \underline{\al}=  \hat{\cal O}\hat{B} \de \underline{\hat\al} \;, \qquad   \hat{\cal O} ={\cal O}  U\,.
\label{41}
\eeq
Finally, we can find the matrix of anomalous dimensions $\Gamma$,
\beq
\Gamma = \left(\begin{array}{cc} \gamma_{{\cal O}_1}&0\\[2mm] 0&\gamma_{{\cal O}_2}
\end{array}\right) \;,
\eeq
 for the operators $ {\cal O}  $, or to be more exact, 
for two linear combinations in $\hat{\cal O}$. To this end we differentiate both sides in (\ref{41}) over $\partial L$ and arrive at
\beq
0=- {\cal {\hat O}} \Gamma \hat{B}\de \underline{\hat\al} + {\cal {\hat O}} \hat{B} \hat{B}\de \underline{\hat\al}  \;,
\label{43}
\eeq
implying, in turn, that 
\beq
\Gamma =\hat{B}\,,
\label{44}
\eeq
cf. Eq. (\ref{35}).
In deriving (\ref{43}) we used Eq. (\ref{38}).
From the above we conclude that
\beq
 \langle {\cal {\hat O}}_1(x)\,{\cal {\hat O}}_1(0) \rangle\propto \frac{1}{(x^2)^4} \, \frac{1}{(x^2\mu^2)^{\lambda_-}} \;,
\eeq
where $\lambda_-$ is the lowest eigenvalue of the matrix $\hat B$ and, hence, of the matrix $B_*$, see Eq. 
(\ref{eq:L}).\footnote{In fact, tracking the $\mu$ dependence in the prefactors one would be able to see that $\mu$ in the denominator 
will be replaced by $\Lambda_-$, in much  the same way as in passing from Eq. (\ref{fourteen}) to (\ref{16p}).}
Following \cite{AGJ} to simplify notation we will denote
\beq
\lambda_- = \beta^\prime_{\rm mag} \;.
\eeq

\vspace{2mm}

Summarizing this section we conclude that
\begin{equation}
\vev{\TEMT(x) \TEMTb(0)}_{\ma} \propto  \frac{1}{(x^2)^{4 + \bestp|_{\ma}}} \,. 
\end{equation}
Since the Seiberg dual correlators must coincide in the corresponding IR fixed points,
\begin{equation}
\vev{\TEMT(x) \TEMTb(0)}_{\ma} \overset{\rm{\scriptscriptstyle IR}}{\longleftrightarrow} \vev{\TEMT(x) \TEMTb(0)}_{\el} \;,
\end{equation}
 we confirm that $\left. \beta_*^\prime\right|_{\rm el} =  \left. \beta_*^\prime\right|_{\rm mag}$ holds as stated in \EQ\eqref{three} in the introduction. This is the central result of our paper.

\section{Relation between \boldmath{$\gastp$}  and  \boldmath{$\bestp $} }
\label{sec5}

The $\be$ function of the electric theory \eqref{eight}  is essentially a relation between the 
matter-field anomalous dimension $\ga_Q$ and the $\be$ function itself.  Since we have gained information 
on $\bestp$ we may exploit this fact to deduce information on  $(\ga_Q')_*$ by directly differentiating at the IR fixed point
\beq
\label{eq:useful}
\beta^\prime_* = -\frac{\alpha^2_*}{2\pi}\,\,  \frac{N_f}{1- \frac{\alpha_*}{2\pi}N_c} \,\left(\gamma^\prime_{Q}\right)_* \;.
\eeq
Hence, $\bestp$ and $ \gastp$ are proportional to each other throughout the CW with a coefficient depending on the unknown critical coupling $\al_*$. 
We note that the relation \eqref{eq:useful} is generally scheme-dependent and so is the fixed point coupling $\al_*$ but $\bestp$ and $\gastp$ are scheme-independent under analytic redefinitions of the coupling, cf. footnote \ref{foot:Wein}.
The fact that both $\bestp$ and $\gastp$ are zero simultaneously might not be completely accidental.  
They both describe the perturbation around the fixed point for a gauge theory with massive mater, e.g. \cite{DelDebbio:2013qta}.
We remind the reader that $\ga_Q$ equals minus the anomalous dimension of the mass 
to all orders in perturbation theory  in ${\cal N}=1$ supersymmetry. 

With \eqref{three} one obtains a strong coupling relation but unfortunately the right hand side 
contains the two unknowns $\alpha_{*}$ and $\gastp$ for which we cannot solve simultaneously.
Nevertheless, one can deduce interesting  information. 
Since, $\bestp$ as a function of  $N_f$ starts at zero for $N_f = 3 N_c$ and then  raises and lowers towards 
 zero  again at $N_f = \frac{3}{2} N_c$, there must be at least two number of flavours,
 $N_f^{(w)} > N_f^{(s)}$, for which 
 \begin{equation}
 \bestp|_{N_f^{(w)}}  =  \bestp|_{N_f^{(s)}} \;,
 \end{equation}
 holds (the superscripts $w$ and $s$ stand for weak and strongly coupled with regards to the electric coupling). 
  As we expect the electric coupling to become continuously 
 stronger towards the lower edge  of the CW  one finds 
 \begin{equation}
 \al^*|_{N_f^{(w)}} < \al^*|_{N_f^{(s)}} \quad  \Leftrightarrow \quad \gastp|_{N_f^{(w)}} > \gastp|_{N_f^{(s)}} \;,
 \end{equation}
the curious result that  $\gastp$ is larger when the theory is weakly coupled and vice versa.

\section{Near the edges of the conformal window}
\label{edgess}

With the knowledge of the $\be$ functions one my investigate them at weak coupling. 
This is particularly interesting in the magnetic case where there are two  couplings. 

As a warm up we will first consider the electric case.   The $\be$ function is given in \eqref{three} 
and since the electric theory is weakly coupled for $N_f$ just below $3N_c$ we expand in the following 
quantity 
\begin{equation}
\eps \equiv \frac{3N_c - N_f}{N_f} \ll 1 \;,
\end{equation}
for which we find the critical coupling and the slope to be
\begin{equation}
\al_* = \frac{2 \pi}{N_c} \eps    \;, \quad  \bestp =  3 \eps^2  \;,
\end{equation}
upon using \EQ(\ref{9}) for $\ga_Q $.

In the magnetic case we first need to obtain the explicit form of the Yukawa $\be$ function in 
\eqref{eq:beM}. We need the anomalous dimension at leading order in the couplings. 
We have computed them explicitly
\begin{equation}
\label{27p}
\ga_q   = - \frac{\alD}{\pi}  \frac{N_D^2-1}{2 N_D}+  \frac{\alf}{2 \pi} N_f   \;, \quad 
\ga_M = \frac{\alf}{2\pi} N_D   \;,  \quad 
\end{equation}
and find agreement with the results found in section 8 of \cite{KSV}.  
The gauge coupling part is identical to the electric case with the replacement $N_c\to N_D$ and the part proportional 
to the Yukawa coupling $\alf$ is related to the computation in the 
Wess-Zumino model.\footnote{ 
In the Wess-Zumino model with superpotential $W(\Phi) = \frac{Y}{6} \Phi^3$, 
 the $Z$-factor of the superfield $\Phi$, as given in   \EQ 2.7  in \cite{Jack:2005ij} for example,  
 is related to  $Z_q = Z_{\Phi}|_{Y^2 \to 2 {\rm f}^2 N_f }$. Moreover, $Z_M = Z_q|_{N_f \to N_D, g_D \to 0}$. 
 Our explicit computation passes this crosscheck.} 
  Assembling \eqref{eq:beM} and (\ref{27p})  we get the explicit Yukawa $\be$ function to leading order,
 \begin{alignat}{2}
 \label{eq:be2}
 &  \be_{\rm f}  &\;=\;&  \alf \,\left[ \frac{\alf}{2 \pi} ( N_D+2 N_f) - \frac{2 \alD}{\pi}  \frac{N_D^2-1}{2N_D}\right]  \;.
 \end{alignat}
  \begin{centering}
\begin{figure}[h!]
\begin{center}
\includegraphics[width=0.6\linewidth]{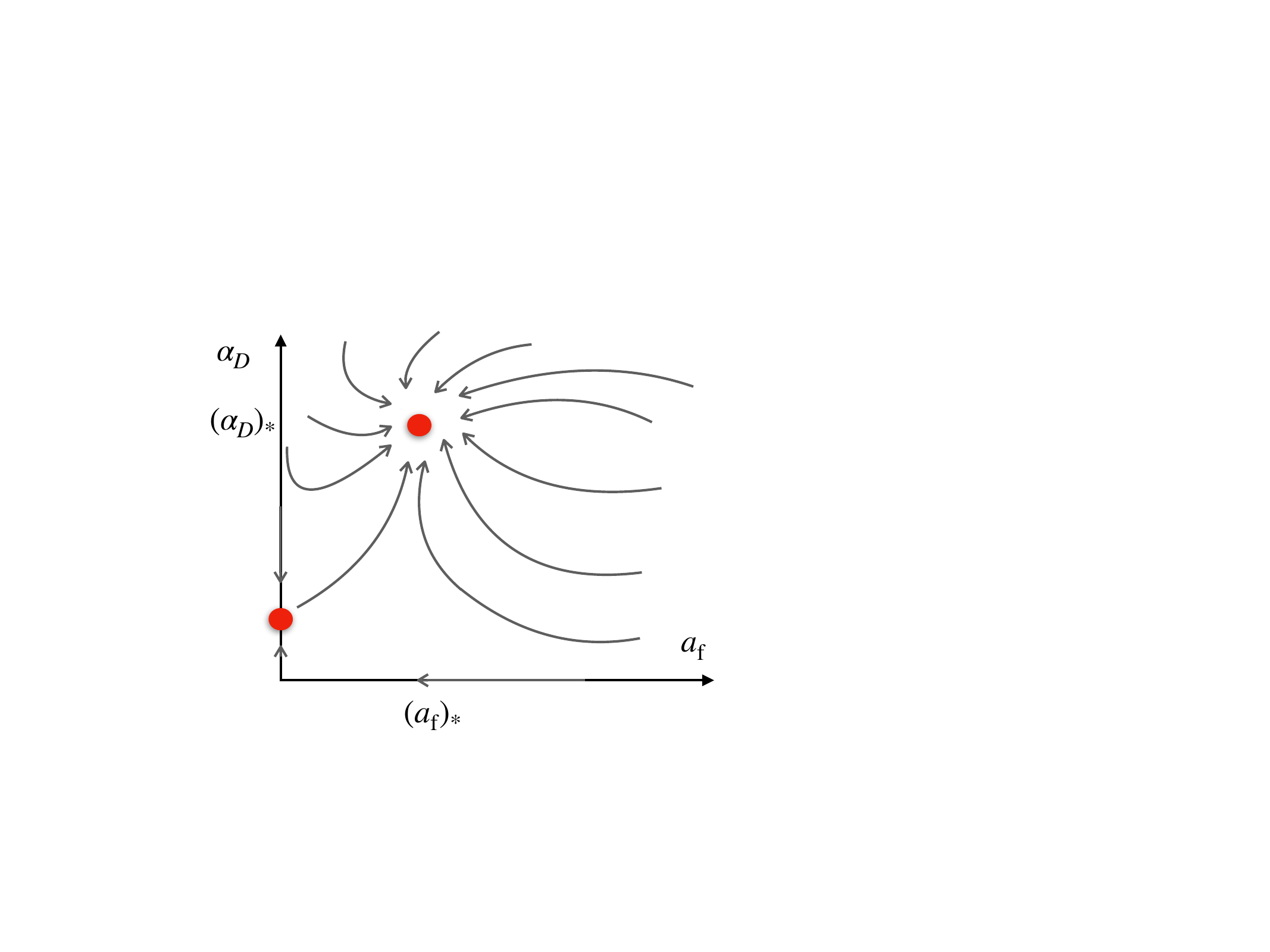}
	\caption{\small Illustration of the RG flow in the  magnetic theory. Red dots denote IR fixed points given in 
	\eqref{eq:magIR} for $N_f$ just above $\frac{3}{2} N_c$ where the magnetic theory is weakly coupled. 
		}
	\label{fig:flowmag}
	\end{center}
\end{figure}
\end{centering}
 The function $\be_{\rm f}$ differs from \EQ (64) in  \cite{KSV} by a simple typo, namely, in \cite{KSV} one should replace 
$$ \frac{ \alD}{\pi}  \to   \frac{ 2\alD}{\pi} \,, $$ cf.  our equation (\ref{eq:be2}). This typo seems to have propagated further in 
 their analysis and we thereby correct the IR fixed point found in that paper. 
In analogy to the electric case we define
\begin{equation}
\eps_D = \frac{3 N_D-N_f}{N_f} \ll 1 \;,
\end{equation}
for the dual magnetic theory to find the fixed point  for the $\be$ functions given in \eqref{eq:be2} with $\be_{\rm f}$ 
approximated as above.  Assuming an ansatz of the form $\alD, \alf \propto \eps_D$
we find the following  two solutions
\begin{equation}
\label{eq:magIR}
\frac{N_D}{ {2} \pi}(\alD ,  \alf)_* =  \eps_D \left\{\begin{array}{ll} 
(1,0)   &  \alf =0    \\[0.1cm]
(7  ,2  )   & \alf \neq 0
  \end{array} \right. \;.
\end{equation} 
The first fixed point with no Yukawa is of the   Banks-Zaks type  whereas the second one with the Yukawa 
coupling switched on is less well-known.  The Banks-Zaks fixed point is unstable as the RG flow tends to the other fixed point for $\alf \neq 0$ (cf. \FIG\ref{fig:flowmag} and \cite{KSV}).

In order to obtain the slope we need the eigenvalues of the $B_*$-matrix  \eqref{eq:L},
 for which we find
\begin{equation}
\label{eq:Bmi}
\lamin  =  21  \,\eps_D^2    \;, \quad \lamax = 14  \, \eps_D   \;,
\end{equation}
such that $\lamin < \lamax$ for $\eps_D \ll 1$. 
Since the slope of the $\be$ function is determined by the minimal eigenvalue we finally get the slope in terms of $\eps_D$\footnote{The result in \eqref{eq:bestp} can be compared to the one in 
 \cite{AGJ} where they obtained $\best'|_{\ma} =\frac{21}{4}  \eps_D^2$, upon using the conversion $\sig \equiv  \frac{3}{2}-\frac{N_f}{N_c} = \frac{3}{4} \eps_D$, 
 which differs by a factor of $4$.}
\begin{equation}
\label{eq:bestp}
 \best'|_{\el} =  \best'|_{\ma} =21  \,\eps_D^2 \;.
\end{equation}
 
It is also interesting to consider the eigenvectors is this approximation. We find the following 
non-orthogonal eigenvectors 
 \begin{equation}
 (\vec{v}_-)^T  =  (1, \tfrac{63}{2} \eps_D)  \;,  \quad 
 (\vec{v}_+)^T  =   \frac{1}{\sqrt{53}}(2 ,- 7)  \;,
 \end{equation}
corresponding to the eigenvalues given above. We infer that for small $\eps_D$ 
 the gluonic operator dominates over the Yukawa term.  

\section{Conclusions}    
\label{sec:conc}

In this paper we have shown  that  the slopes of the  $\be$ function at the IR fixed point,
are equal to each other  \eqref{three} in the electric and the magnetic theories of the Seiberg duality.
This result was derived some time  ago using the Konishi currents and the Kutasov construction \cite{AGJ}. 
We found a simpler way to obtain this result by matching the  two-point function
 of the trace of the energy momentum tensor in the electric and magnetic theory.  By the very assumption of
 the Seiberg duality such a geometric quantity has to match and since its scaling is governed by 
 $\bestp$ the result follows. 
 
 In passing we obtained   
a new relation between between  $\gastp$ and  $\bestp $ given in \EQ\eqref{eq:useful}. 
The RG flow near the edge of the conformal window previously discussed is \cite{KSV} is corrected.
We obtain in addition the corresponding eigenvalues and eigenvectors  of the flow in the magnetic theory. 
These results might be useful in that $\bestp $ and $\gastp$  are the quantities that 
describe perturbations around a fixed point in a gauge theory with matter. 

\paragraph{Acknowledgments:} 

The work of MS  is supported in part by DOE grant DE-SC0011842. RZ is supported by a CERN associateship and an STFC Consolidated Grant, ST/P0000630/1. 
MS thanks Andrey Johansen for multiple conversations.
RZ is grateful to Steve Abel, Ken Intriligator  and Thomas Ryttov for correspondence and or discussions. 

\appendix

\section{The trace of the energy momentum tensor}
\label{app:defs}

\setcounter{equation}{0}
 \renewcommand{\theequation}{\thesection.\arabic{equation}}

In superfields, the anomalies in the TEMT and in the divergence of the $R$ current are given by a unified formula for the hypercurrent ${\mathcal J}_{\alpha\dot\alpha}$ (see the second reference in \cite{{Shifman:1999mk}}, \SEC10.27.4)
which for our choice of the superpotential (\ref{eqn14}) takes on the form
\begin{align}
\partial^{\alpha\dot{\alpha}}\mathcal{J}_{\alpha\dot{\alpha}}&=-\frac{i}{3}D^{2}\left\{\left[-\left(\frac{\gamma_{M}}{2}+\gamma_q\right)
\mathcal{W}\right]\right.\nonumber\\
&\qquad\qquad\quad-
\left.\frac{1}{16\pi^{2}}\left[3N_D -N_f  + N_f \gamma_{q})\right]
\mathrm{Tr}\, W^{2}\right\}+\mathrm{H.c.},
\label{eqn49}
\end{align}
where $D$ is the spinorial derivative which singles out the $\theta^2$ component on the right-hand side. Eq. (\ref{eqn49}) refers to the magnetic theory.
In the electric theory ${\mathcal W}=0$, so the first line disappears, $N_D \to N_c$ and $\gamma_q\to \gamma_Q$. Equation (\ref{eqn49}) implies
\beq
\label{eq:TEMT}
\theta^\rho_{\,\rho} = \CG  \beta(\alpha_D, a_{\rm f})  \OG + \CW \beta_{\rm f}(\alpha_D, \alf)  \OW  \;,
\eeq
where 
\beq
\CG  =  \left(16\pi\,\alpha^2 \right)^{-1} \,,\qquad  \CW =\sqrt{\frac{4\pi}{\alf}} \;,
\,
\label{a3}
\eeq
and
\beqn
\label{a4}
\OG  \!\!\!&=&\! \!\! - 2 {\rm Re} \left[W^{\alpha\,a} W^a_\alpha
\right]_{\theta^2{\rm mag} } =
\left[G_{\mu\nu}^a G^{\mu\nu\, a}-2D^2 -4i\bar{\lambda}_{\dot\alpha}^a {\mathcal D}^{\dot\alpha\alpha}\lambda_{\alpha}^a
\right]_\ma \;, \\[2mm]
\OW  &=& -2 {\rm Re} \, \left[M^i_{{j}} q_i \tilde{q}^{\, j}\right]_{\theta^2} = -2 {\rm Re} \, \left[ q F_M  \tilde{q}
+ \psi\tilde{\psi}M
+{\rm perm}\right]  \;.
\label{a5}
\eeqn
Here $D$ is the $D$-term of the gauge superfield and $F$ is the $F$-term of the chiral superfields.

\section{The $R$ current  and Konishi current correlators }
\label{app:KR}

Continuing from \SEC\ref{sec:electric} we can find the scaling dimension of the $R$ and Konishi current 
in the electric theory 
without much further effort. 
The (unimproved) $R$ current, enters the same supermultiplets as the energy-momentum tensor,
\begin{equation}
R_{\mu}=-\frac{1}{g^{2}}\lambda^{a}\sigma_{\mu}\bar{\lambda}^{a}+\frac{1}{3}\sum_{f}\left(\psi_{f}\sigma_{\mu}\bar{\psi}_{f}-2i\phi_{f}\overset{\leftrightarrow}{D}_{\mu}\bar{\phi}_{f}\right).
\label{eqn59.20}
\end{equation}
This (unimproved) current is not conserved  because of the chiral anomaly. The $R$ symmetry is anomalous,
\beq
\pt_\mu R^\mu = \left[  \left( -24\pi\,\alpha^2 \right)^{-1} \beta(\alpha) \, \left(G\tilde G+...\right)\right]_{\mu} \;,
\eeq
{where $G\tilde G  \equiv G_{\mu\nu}^a \tilde{G}^{\mu\nu\, a}$ with the dual tensor 
 $\tilde{G}^{\mu\nu\, a}=\frac 1 2 \varepsilon^{\mu\nu\alpha\beta} {G}^{\alpha\beta\, a}$} and  $$G_{\mu\nu}^a \tilde{G}^{\mu\nu\, a} +...\propto {\rm Im} W^2\,,
 $$
 cf. (\ref{a4}). Taking into account the fact that the anomalous dimension of $ {\rm Im} W^2$ is the same as that of $ {\rm Re} W^2$
we can readily calculate the two-point function 
\beq
 \langle \pt_\mu R^\mu (x) \pt_\nu R^\nu(0) \rangle \propto   \left[\left.\left(-24\pi\,\alpha^2 \right)^{-1} \beta(\alpha)\right|_\mu\right]^2  \frac{1}{(x^2\Lambda^2)^{\beta^\prime_*}}\,,
\label{19p}
\eeq
from the anomaly in the hypercurrent which includes both operators $R_\mu$ and $\theta_{\mu\nu}$ \cite{Shifman:1999mk}.

The result is the same as in (\ref{fourteen}), with the replacement $G^2\to G\tilde G$. Then, we can drop the derivatives in (\ref{19p}) to obtain
\beq
\langle R_\mu (x) R_\nu(0) \rangle \propto \frac{1}{(x^2 )^{ 3}} \, \frac{1}{(x^2\Lambda^2)^{\beta^\prime_*}}\,.
\label{20p}
\eeq
This $x$ scaling law differs from that in (\ref{16p}) by the engineering dimension of $R_\mu$, namely $d_R = 3$ vs $d_\theta=4$.
The anomalous dimensions are exactly the same {as they have to be since they belong in the same supermulriplet.}

\vspace{3mm}

Finally, let us consider the two-point function of the Konishi current. There is a small nuance here which deserves to be discussed.
The  Konishi current is defined as
\beq
K_\mu=\sum_{\psi_f, {\tilde\psi^f}}\left(-\psi_f \sigma_\mu\bar\psi^f - \phi_f i\stackrel{\leftrightarrow}{\mathcal D} _\mu\bar\phi^f
\right)\,.
\label{21}
\eeq
By the same token,
the flavor-singlet Konishi current is {\em not} conserved due to the the anomaly,\,\footnote{In the superfield language $\bar{D}^2{\mathcal J}_K =\frac{N_f}{2\pi^2} {\rm Tr} W^2$.    The relation between   $K_\mu$ in (\ref{21}) and  
${\mathcal J}_K$ is as follows:  $K_\mu$ is the $\theta\bar\theta $ component of  ${\mathcal J}_K$.   }  
\beq
\partial^\mu K_\mu = \frac{1}{48\pi^2} N_f G\tilde{G}\,.
\eeq
Next, we note that  the operator $G\tilde{G}$ on the right-hand side resides in the same superfield $W^2$ as $G^2$.
Therefore, the anomalous dimension of  $G\tilde{G}$ is  the same as that of $G^2$,
\beq
\left(\gamma_{G\tilde G}\right)_* = \left(\gamma_{G^2}\right)_* \,,
\label{23}
\eeq 
where the latter has already been given in \eqref{thirteen}.
Is there a difference compared to the cases of TEMT and $R_\mu$? 

\vspace{1mm}

The answer is  positive. Indeed, 
$ \theta^\rho_\rho$ has the zero anomalous dimension.  This is the reason why Eq. ({\ref{16p}) has no sliding scale $\mu$. 
The cancellation of $\mu$ is achieved thanks to the prefactor defined above Eq. (\ref{15p}).
At the same time, $K_\mu$ has a nonvanishing anomalous dimension. Hence,  as a result, the sliding scale $\mu$ is present in the correlation function
\beq
\langle K_\mu (x) K_\nu(0) \rangle \propto \frac{1}{(x^2 )^3} \, \frac{1}{(x^2\mu^2)^{\beta^\prime_*}}\,.
\label{24}
\eeq
If we compare with Eq. (\ref{20p}) we will see the sliding $\mu^2$ instead of fixed $\Lambda^2$ -- this is the only difference. The  IR scaling law (\ref{24}) for the Konishi current was first derived in \cite{AGJ}.
%
%
%
%

%
%
%
%
\end{document}